\documentclass[aps,pra,twocolumn,floatfix,amsmath]{revtex4}

\usepackage{pgfplots}
\pgfplotsset{compat=1.13}
\usepackage[T1]{fontenc}
\newcommand{\neurons}[0]{$50000$}
\newcommand{\colorbar}[2]{
\begin{tikzpicture}
\begin{axis}[
    hide axis,
    scale only axis,
    height=0pt,
    width=0pt,
    colormap/blackwhite,
    colorbar horizontal,
    point meta min=#1,
    point meta max=#2,
    colorbar style={
    	rotate=90,
        width=0.12\textwidth,
        xtick={#1,0,#2},
        pos={5cm},
        xticklabel style={anchor=east},
        xticklabel pos=right,
    }
    ]
    \addplot [draw=none] coordinates {(0,0) (1,1)};
\end{axis}
\end{tikzpicture}
}
\newcommand{\colorbarh}[2]{
\begin{tikzpicture}
\begin{axis}[
    hide axis,
    scale only axis,
    height=0pt,
    width=0pt,
    colormap/blackwhite,
    colorbar horizontal,
    point meta min=#1,
    point meta max=#2,
    colorbar style={
        width=0.4\textwidth,
    }]
    \addplot [draw=none] coordinates {(0,0)};
\end{axis}
\end{tikzpicture}
}
\newcommand{\errors}[2]{
\begin{tabular}{p{2.7cm}p{2.6cm}p{2.5cm}}
&$E_\textup{rms} = #1\%$&$E_\textup{rms} = #2\%$
\end{tabular}
}

\usepackage{graphicx}

\newcommand{\scalebarbackground}[5][white]{
 \begin{tikzpicture}
  \draw (0,0) node[anchor=south west,inner sep=0] (image) { #2 };
  \begin{scope}[x={(image.south east)},y={(image.north west)}]
   \fill [fill=black, fill opacity=0.5] (0.04,1.3em) rectangle (#5*#4/#3+0.04,0.1em);
   \draw [#1, line width=0.2em] (0.04,1.2em) -- node[below=1pt,inner sep=0.1em, font=\footnotesize] {\SI{#5}{\nano \meter}} (#5*#4/#3+0.04,1.2em);
  \end{scope}
 \end{tikzpicture}
}

\renewcommand{\v}[1]{\mathbf{#1}}
\usepackage{siunitx}
\usepackage{mathrsfs}

\begin{document}

\title{Propagation based phase retrieval of simulated intensity measurements using artificial neural networks}

\author{Z.~D.~C.~Kemp}
\email[]{zachary.kemp@monash.edu}
\affiliation{Monash School of Physics and Astronomy, Monash University, Victoria 3800, Australia}



\begin{abstract}
Determining the phase of a wave from intensity measurements has many applications in fields such as electron microscopy, visible light optics, and medical imaging. Propagation based phase retrieval, where the phase is obtained from defocused images, has shown significant promise. There are, however, limitations in the accuracy of the retrieved phase arising from such methods. Sources of error include shot noise, image misalignment, and diffraction artifacts. We explore the use of artificial neural networks (ANNs) to improve the accuracy of propagation based phase retrieval algorithms applied to simulated intensity measurements. We employ a phase retrieval algorithm based on the transport-of-intensity equation to obtain the phase from simulated micrographs of procedurally generated specimens. We then train an ANN with pairs of retrieved and exact phases, and use the trained ANN to process a test set of retrieved phase maps. The total error in the phase is significantly reduced using this method. We also discuss a variety of potential extensions to this work.
\end{abstract}

\maketitle
Propagation based phase retrieval reconstructs the exit phase of a specimen from out-of-focus intensity images. It is applicable to many optical modalities, with phase retrieval techniques having been used in visible light~\cite{barty1998quantitative}, x-ray~\cite{mayo2003x}, electron~\cite{volkov2002new}, and neutron~\cite{allman2000imaging} optics, and can be achieved using a variety of techniques~\cite{paganin1998noninterferometric, oxley, gerchberg, rivenson2}. Linear phase retrieval algorithms have significant performance advantages over iterative methods, but often come with disadvantages such as poor noise-stability~\cite{GUREYEV200453}.

Artificial neural networks (ANNs) have found application in fields such as natural language processing~\cite{socher2012deep}, game playing~\cite{silver2016mastering}, and medical imaging~\cite{bar2015deep}. Here we employ an ANN to improve the accuracy of a propagation-based phase retrieval algorithm based on the transport-of-intensity equation (TIE)~\cite{teague1983deterministic}.

The technique described here utilizes a set of training data comprising pairs of exact and retrieved phases obtained from specimen simulations which are procedurally generated (i.e., they are generated algorithmically using randomized parameters). The training data is used to iteratively optimize the model, which is then applied to a test set of retrieved phases, resulting in output phases of significantly improved accuracy.

ANNs have been employed to solve one dimensional~\cite{burian2003recurrent} and two dimensional~\cite{burian2000two} phase retrieval problems. That work focused on the high parallelizability of neural networks and the consequent fast convergence properties relative to conventional iterative approaches. More recently, ANNs have been employed to analyze convergent beam electron diffraction patterns~\cite{pennington2014third,xu2017deep}. Our method utilizes an ANN to improve the accuracy of phase retrieval algorithms. This work does not require any \textit{a priori} knowledge of phase retrieval, other than for the purposes of generating training data and performing the initial phase recovery, so it is more easily generalizable to a wide variety of phase retrieval techniques. Rivenson \textit{et al.}~\cite{rivenson2017phase} employed an ANN to enable fast and accurate computation of the exit surface wavefield using a single in-line hologram. Their method utilized experimental training data using visible light, with the training pairs consisting of a complex wavefield obtained from a single in-line hologram as the input, and a corresponding wavefield, obtained using an eight image iterative phase retrieval algorithm, as the target for the ANN. This process is labor intensive, and would typically result in a small volume of training data ($100$ training pairs in their work). Our method improves on this by utilizing procedurally generated specimen simulations to produce a large volume of training data. It also enables the scaled projected potential to be used directly as the target phase, resulting in training data whose accuracy is limited only by the precision and resolution used in the simulations.

The TIE relates transverse derivatives of the phase of the exit surface wavefunction to the longitudinal derivative of the intensity~\cite{teague1983deterministic,paganin1998noninterferometric}:
\begin{align}
-k\frac{\partial I_0}{\partial z} = \nabla_\perp\cdot\left(I_0\nabla_\perp\varphi_0\right), 
\end{align}
where $k$ is the wavenumber, $\nabla_\perp$ is the transverse gradient operator, $\frac{\partial I_0}{\partial z}\equiv\left.\frac{\partial I}{\partial z}\right|_{z=0}$ is the longitudinal derivative of the intensity, evaluated at the image plane, and $I_0$ and $\varphi_0$ are the intensity and phase, respectively, at the image plane. This can be solved numerically using a Fourier transform method~\cite{paganin1998noninterferometric,volkov2002new}, giving

\begin{align}
\varphi_0 = \frac{k}{4\pi^2}\mathscr F^{-1}\left\{\frac{\v k}{|\v k|^2}\cdot\mathscr F\left[\frac{1}{I_0}\mathscr F^{-1}\left(\v k\frac{\mathscr F\left(\frac{\partial I_0}{\partial z}\right)}{|\v k|^2}\right)\right]\right\},\label{eq:tie_num}
\end{align}
where $\v k$ is the spatial frequency vector, and $\mathscr F$ and $\mathscr F^{-1}$ are the forward and inverse Fourier transforms, respectively. 

Equation \eqref{eq:tie_num} contains singularities at $I_0 = 0$ and $|\v k| = 0$. To compute $\varphi_0$, some form of regularization must be employed. Tikhonov regularization~\cite{tikhonov1977solutions}---which gives the transformations $I_0\rightarrow (I_0^2 + \delta_\textup{int}^2)/I_0$ and $|\v k|^2\rightarrow (|\v k|^4 + \delta_\textup{tie}^4)/|\v k|^2$, where $\delta_\textup{int}$ and $\delta_\textup{tie}$ are the respective regularization parameters---is used in this work. We use $\delta_\textup{int} = 0.1I_\textup{in}$ and $\delta_\textup{tie} = 0.1 / (am)$, where $I_\textup{in}$ is the incident intensity, and $a$ and $m$ are the width of the image in units of length, and pixels, respectively. The intensity derivative is computed using the finite difference approximation
\begin{align}
\frac{\partial I_0}{\partial z} = \frac{I_+ - I_-}{2\Delta f},\label{eq:fda}
\end{align}
where $I_+$ and $I_-$ are the over- and under-focus images, respectively, and $\Delta f$ is the magnitude of the defocus, which, in this work, is the same for both images.

To assess the accuracy of the retrieved phase, we use a normalized root mean square (rms) error metric given by
\begin{align}
E_\textup{rms} = \frac{\sum_{i,j}{(\varphi^\textup{ex}_{i,j} - \varphi^\textup{ret}_{i,j}})^2}{\sum_{i,j}\left( \varphi^\textup{ex}_{i,j}\right)^2},\label{eq:rmse}
\end{align}
where $\varphi^\textup{ex}$ and $\varphi^\textup{ret}$ are the exact and retrieved phases, respectively. To improve sampling in Fourier space, we apodize the images with a circular window function of radius $0.5a$. Equation \eqref{eq:rmse} is computed over this circular region. To calculate the average rms error for a collection of test phases, we take the mean of Eq.~\eqref{eq:rmse}:

\begin{align}
\bar E_\textup{rms} = \frac{1}{N}\sum_p{E_{\textup{rms}}^{(p)}}\label{eq:rms_mean},
\end{align}
where $N$ is the number of pairs in the test set, and $p$ is the index corresponding to each example in the set. Images of the error $E$ are obtained by taking the difference
\begin{align}
E = \varphi^\textup{ret} - \varphi^\textup{ex}.\label{eq:error}
\end{align}
Again, we compute these errors only over the circular window.

A fully connected feedforward ANN employs a series of affine transforms applied to an input vector, to produce some output. Each transform, consisting of a weight matrix and bias vector, is referred to as a layer. In our case, the input is a retrieved phase map, and the output is the improved approximation $\varphi_\textup{out}$ to the true phase shift, which we refer to as the adjusted phase. The input and output are vectors of length $M$ obtained by flattening the $m\times m$ phase maps. In this work, $m = 128$, resulting in $M=16384$. The ANN has the form
\begin{align}
\varphi_\textup{out} = (\varphi_\textup{in}W^{(1)}+B^{(1)})W^{(2)}+B^{(2)},\label{eq:ann}
\end{align}
where $W^{(q)}$ and $B^{(q)}$ are the weights and biases of the $q^\textup{th}$ layer, respectively. Our hidden layer (layer 1) has \neurons{} nodes. We initialize the weights such that $W^{(1)}$ is a rectangular diagonal unit matrix, with $W^{(2)}$ its transpose, and the biases are zero vectors. To model non-linear effects, an activation function must be applied to the hidden layer. This is a non-linear function that is applied pointwise to the output of the given layer. We use a hyperbolic tangent activation function. The choice of initialization for the weights and biases is reasonable because the ideal output, $\varphi_\textup{ex}$, is quantitatively similar to the input, and this initialization causes the output of the untrained ANN to be similar to its input, with the only difference being a scaling due to the use of the activation function.

The ANN is trained by minimizing a cost function $L$ with respect to $W^{(q)}$ and $B^{(q)}$. Our cost function is
\begin{align}
L = \sum_{i,j}(\varphi^\textup{ex}_{i,j} - \varphi^\textup{ret}_{i,j})^2.
\end{align}
We optimize the model using gradient descent, which minimizes the cost function by altering the weights and biases (grouped here as $\theta$), for each iteration, by~\cite{rumelhart1986learning}
\begin{align}
\Delta\theta = -\eta\frac{\partial L}{\partial\theta},
\end{align}
where the parameter $\eta$ is known as the learning rate. Backpropagation~\cite{lecun2015deep} is used to compute the gradients.

\begin{figure}[htbp]
\centering
{\scalebarbackground{\includegraphics[width=0.3\linewidth]{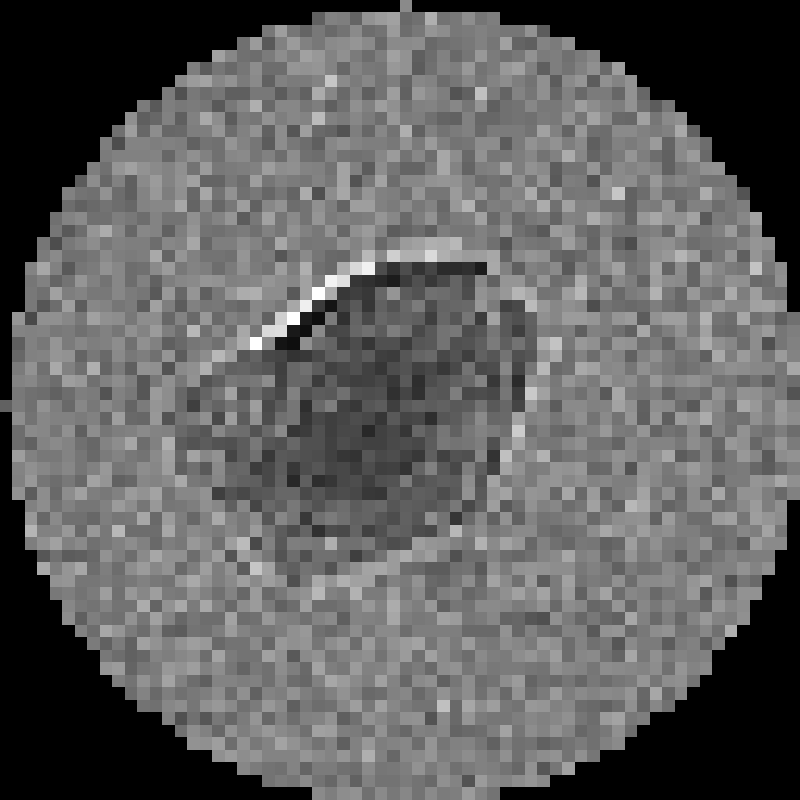}}{150}{1}{50}}{\includegraphics[width=0.3\linewidth]{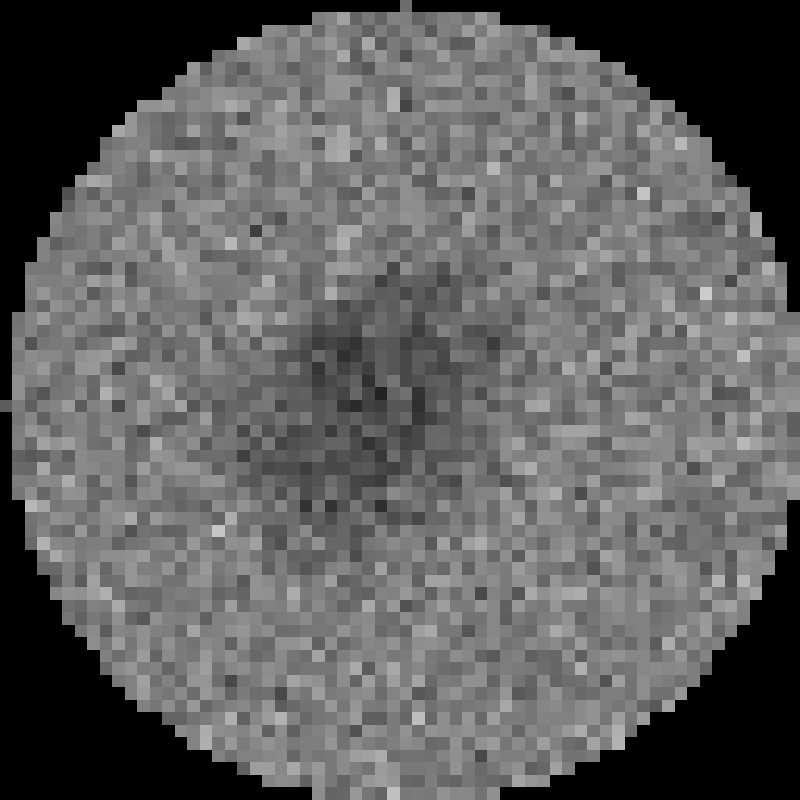}} {\includegraphics[width=0.3\linewidth]{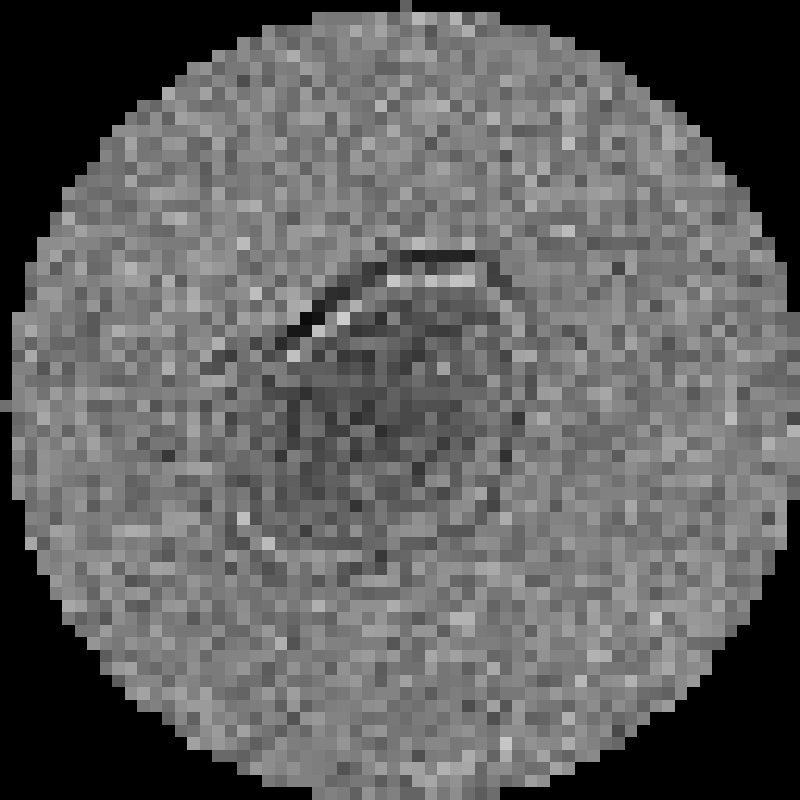}}
\begin{tikzpicture}
\begin{axis}[
    hide axis,
    scale only axis,
    height=0pt,
    width=0pt,
    colormap/blackwhite,
    colorbar horizontal,
    point meta min=0,
    point meta max=2,
    colorbar style={
        width=0.4\textwidth,
        xtick={0,0.5,...,2},
				xticklabels={0,$I_\textup{in}/2$,$I_\textup{in}$,$3I_\textup{in}/2$,$2I_\textup{in}$},
    }]
    \addplot [draw=none] coordinates {(0,0)};
\end{axis}
\end{tikzpicture}
\caption{Example simulated micrographs (under-, in-, and over-focus, from left to right).}
\label{fig:example-training-images}
\end{figure}
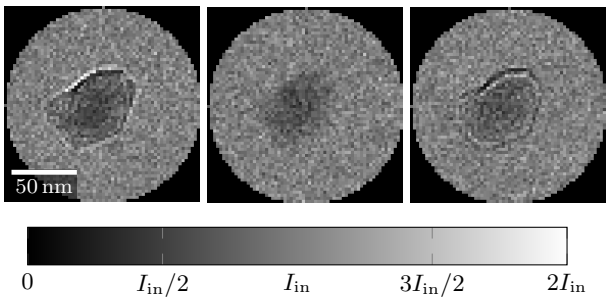

\begin{figure}[tbp]
\centering
{\scalebarbackground{\includegraphics[width=0.3\linewidth]{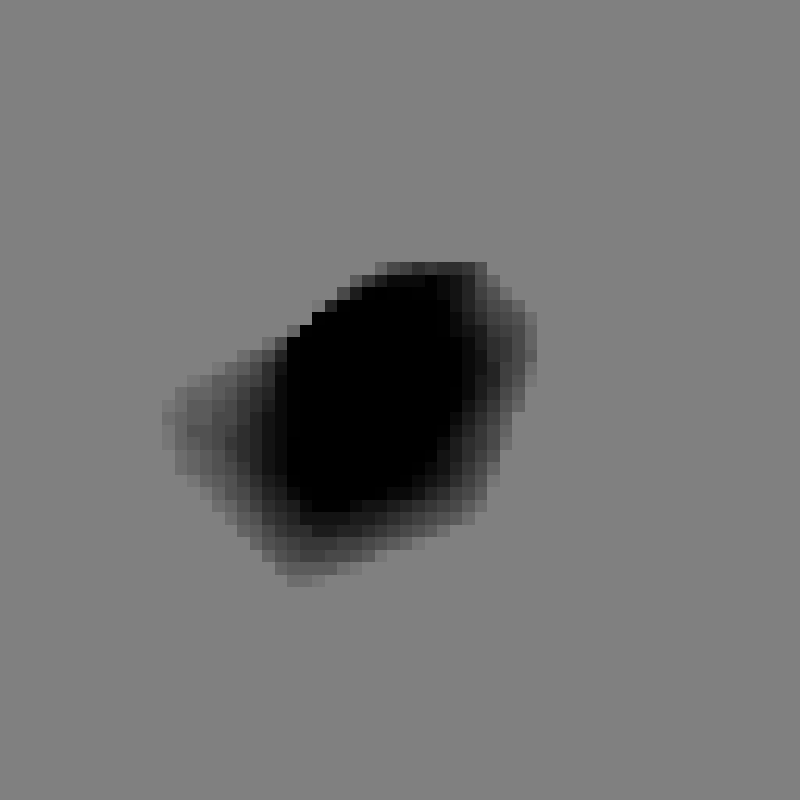}}{150}{1}{50}}{\includegraphics[width=0.3\linewidth]{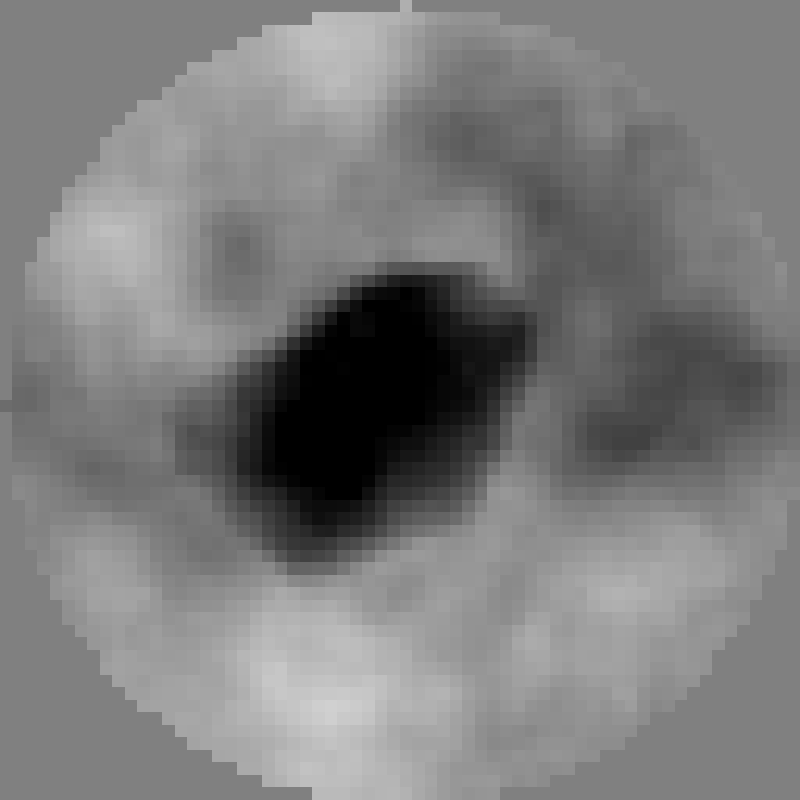}}
\colorbar{-3}{3}
\caption{The pair of exact and retrieved phases, respectively, corresponding to the micrographs shown in Fig.~\protect\ref{fig:example-training-images}.}
\label{fig:example-training-phases}
\end{figure}

For our simulations, we consider phase retrieval in the context of electron microscopy. The ANN training data consists of pairs of exact and retrieved phases. The specimen for each training example is randomly generated using a python script in conjunction with Blender 2.77a~\cite{blender} and its built-in python API. This is achieved by spawning a cube with size, location, and rotation chosen from a uniform pseudo-random distribution, and implementing geometry modifiers (with some of their parameters, again, selected from a uniform pseudo-random distribution) included in the API. The exact phase for each specimen is computed by projecting the electrostatic potential in the electron propagation direction. To simulate attenuation, we include an imaginary part in the potential; specifically, we use a potential of $V_0 = \SI{-17+i}{\volt}$, which is constant throughout the specimen. To compute retrieved phases, under-, in-, and over-focus images are simulated, for $\Delta f =\SI{8}{\micro\metre}$, using a transfer function formalism. We use an image size of $a = \SI{150}{\nano\metre}$ and an electron accelerating voltage of $\SI{300}{\kilo\volt}$. Shot noise at a level of $15\%$ is added to the micrographs for the first training set using the technique described in \cite{kemp2014analysis}. The second training set is generated using the same methods as the first set, with $\Delta f$ increased to $\SI{80}{\micro\metre}$ to exacerbate blurring in the retrieved phases. The micrographs for this set remain noise-free for the purposes of testing the neural network on systematic phase retrieval errors. Equation \eqref{eq:tie_num} is used to retrieve the phase from these images. The simulated micrographs of the first specimen of the first training set are shown in Fig.~\ref{fig:example-training-images}, and the corresponding exact and retrieved phases are shown in Fig.~\ref{fig:example-training-phases}.

Once exact and retrieved phases are prepared, gradient descent is performed using $100$ batches of $50$ pairs each, with $\eta=0.5$, and the optimization process runs for $50$ epochs. Our ANN is implemented using TensorFlow 1.2.1~\cite{tensorflow2015-whitepaper}.

\begin{figure}[htbp]
\centering
{\scalebarbackground{\includegraphics[width=0.3\linewidth]{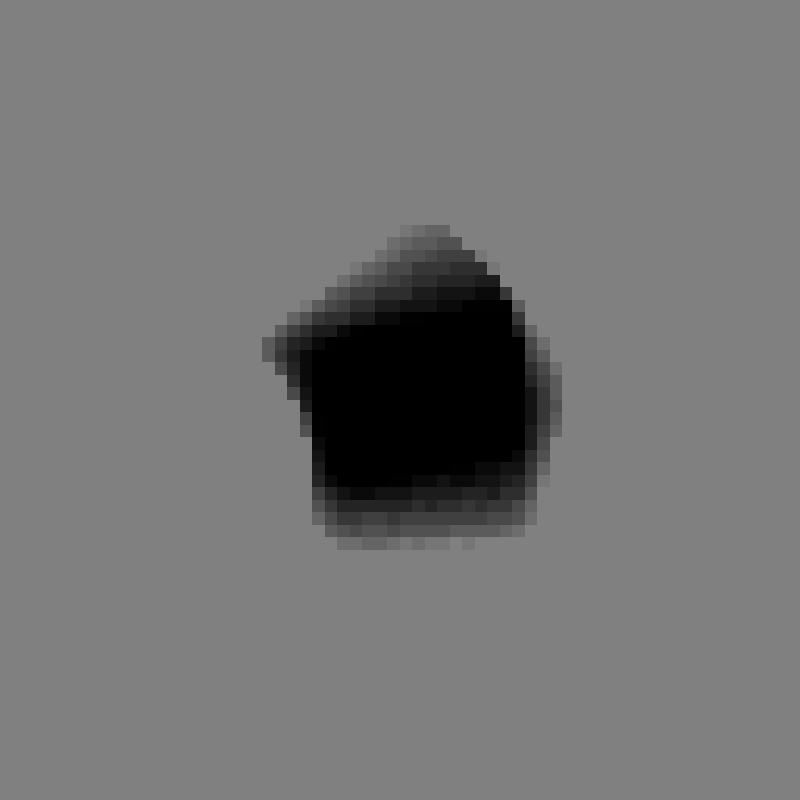}}{150}{1}{50}}{\includegraphics[width=0.3\linewidth]{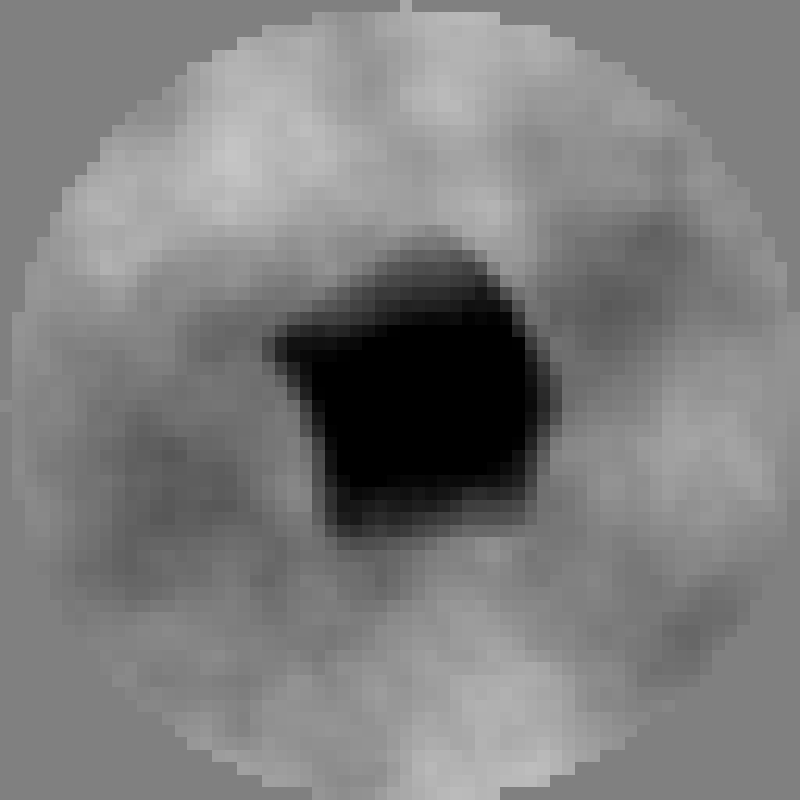}} {\includegraphics[width=0.3\linewidth]{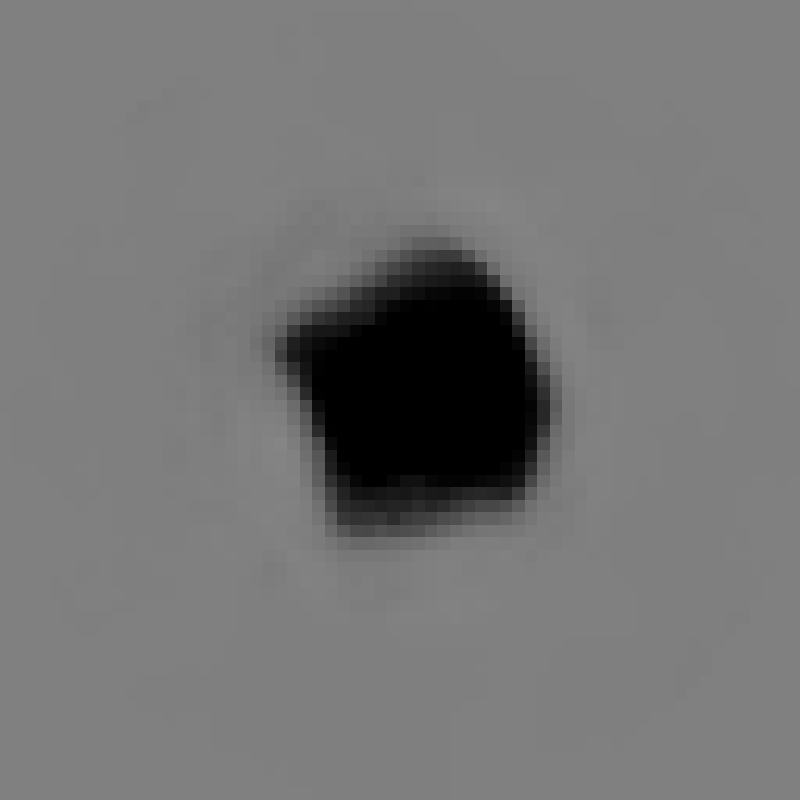}}
\errors{52.3}{14.7}
{\includegraphics[width=0.3\linewidth]{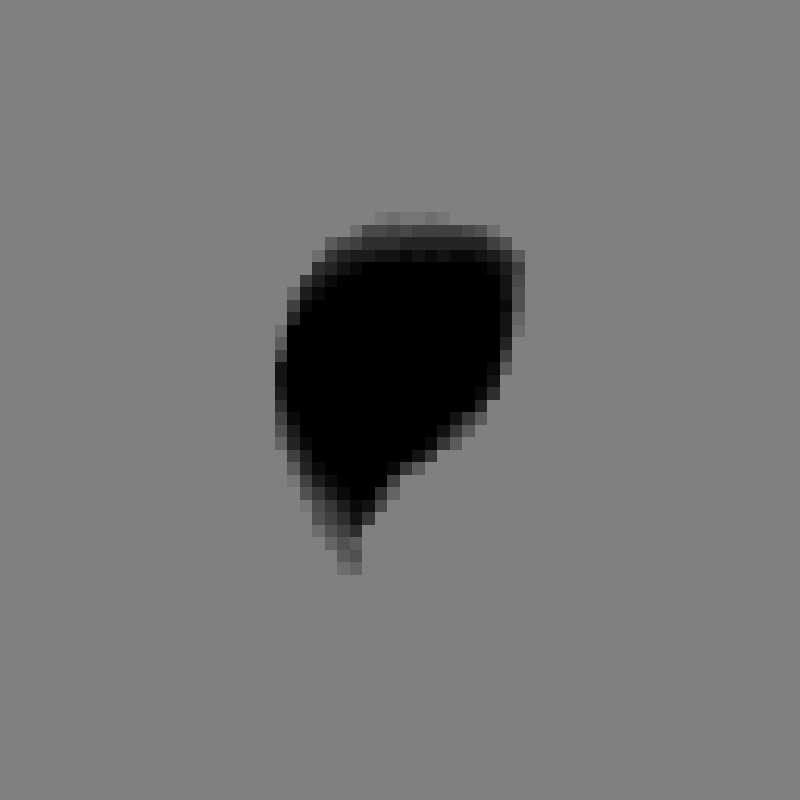}} {\includegraphics[width=0.3\linewidth]{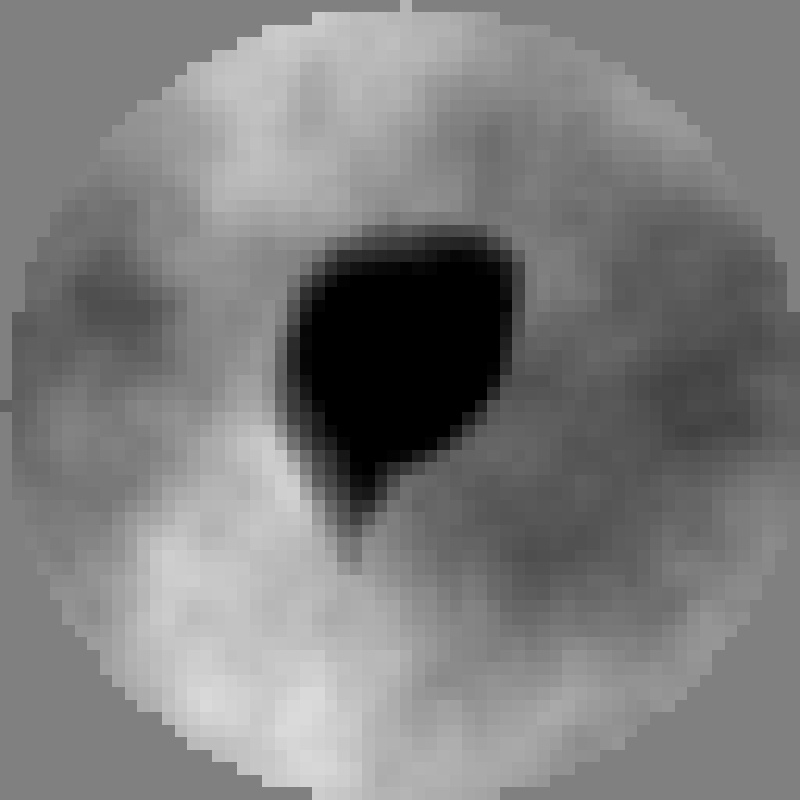}} {\includegraphics[width=0.3\linewidth]{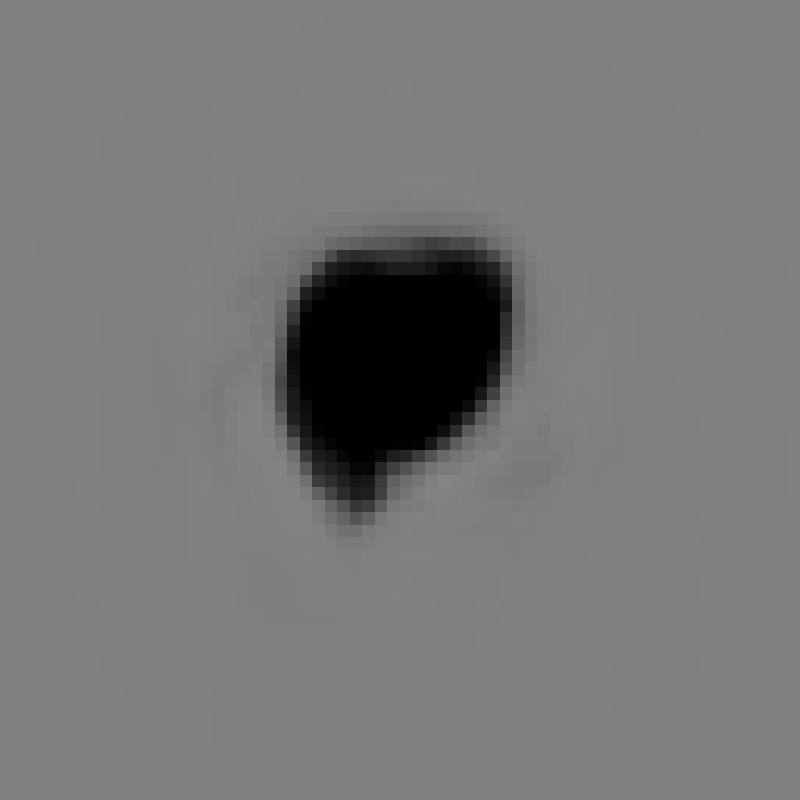}}
\errors{70.6}{23.3}
{\includegraphics[width=0.3\linewidth]{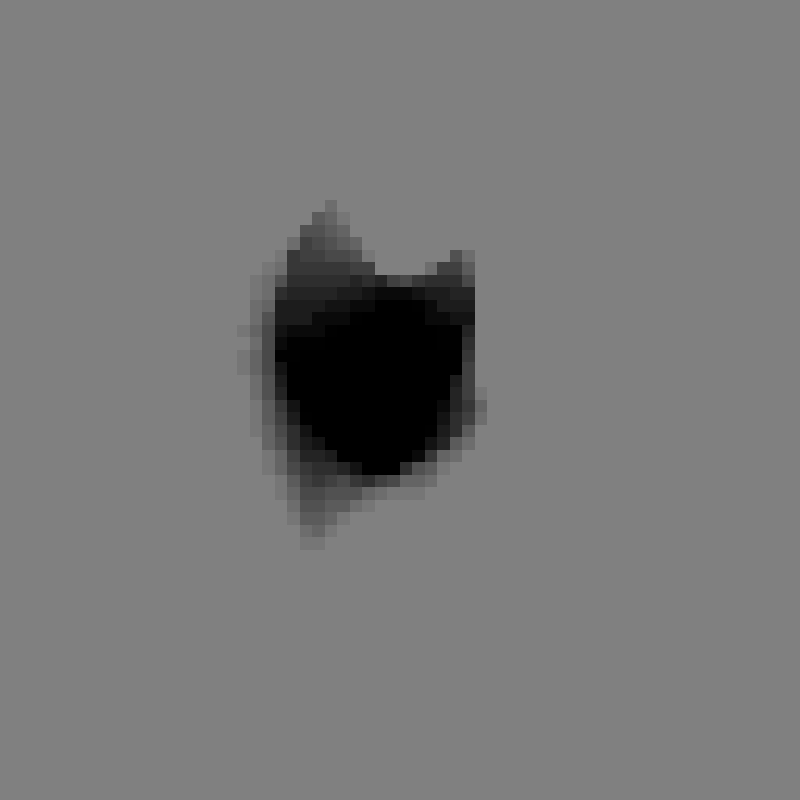}} {\includegraphics[width=0.3\linewidth]{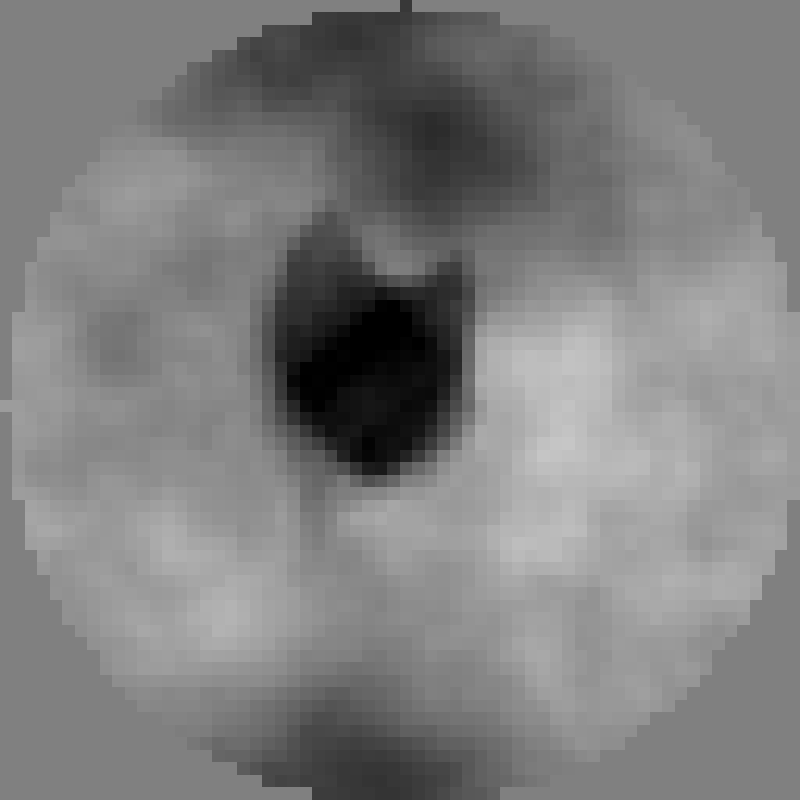}} {\includegraphics[width=0.3\linewidth]{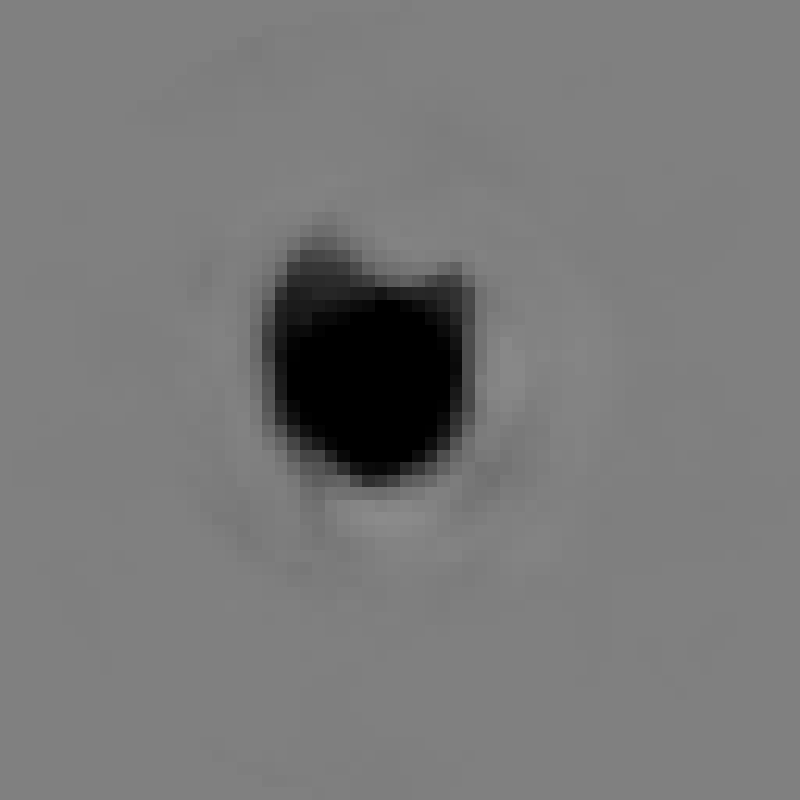}}
\errors{92.2}{15.5}
\colorbarh{-3}{3}
\caption{Exact, retrieved, and adjusted phases, respectively, with the rms error displayed under each retrieved and adjusted phase. Each row shows a procedurally generated specimen from the test set.}
\label{fig:example-test-sets}
\end{figure}

To test the effectiveness of the ANN, we prepared a test set of one hundred exact/retrieved phase pairs, using the same methods as for the training set, again using randomly generated specimens. The ANN trained on the noisy set was applied to the retrieved phases. Results for three example pairs from this test set are shown in Fig.~\ref{fig:example-test-sets}. We calculated the mean rms error over the entire test set using Eq.~\eqref{eq:rms_mean} and found $\bar E_\textup{rms} = 77.8\%$ and $\bar E_\textup{rms} = 19.6\%$ for retrieved and adjusted phases, respectively.

\begin{figure}[htbp]
\centering
{\scalebarbackground{\includegraphics[width=0.3\linewidth]{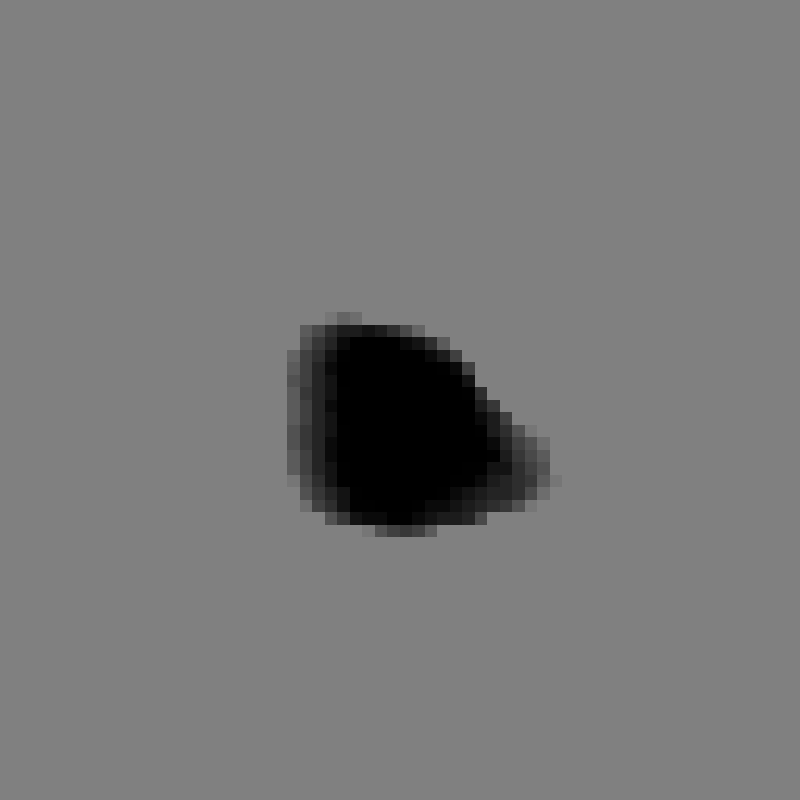}}{150}{1}{50}}{\includegraphics[width=0.3\linewidth]{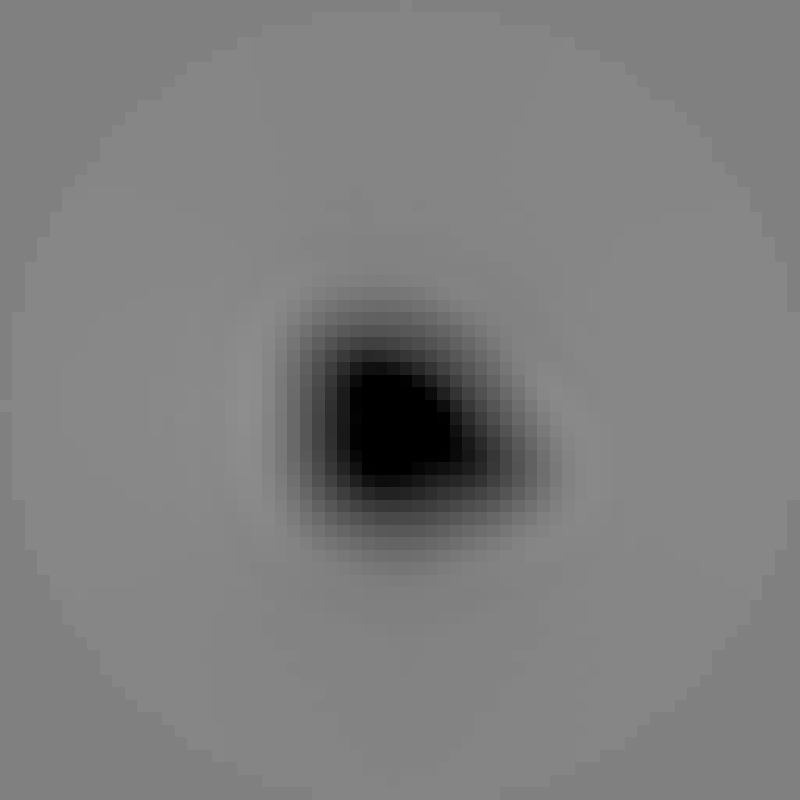}} {\includegraphics[width=0.3\linewidth]{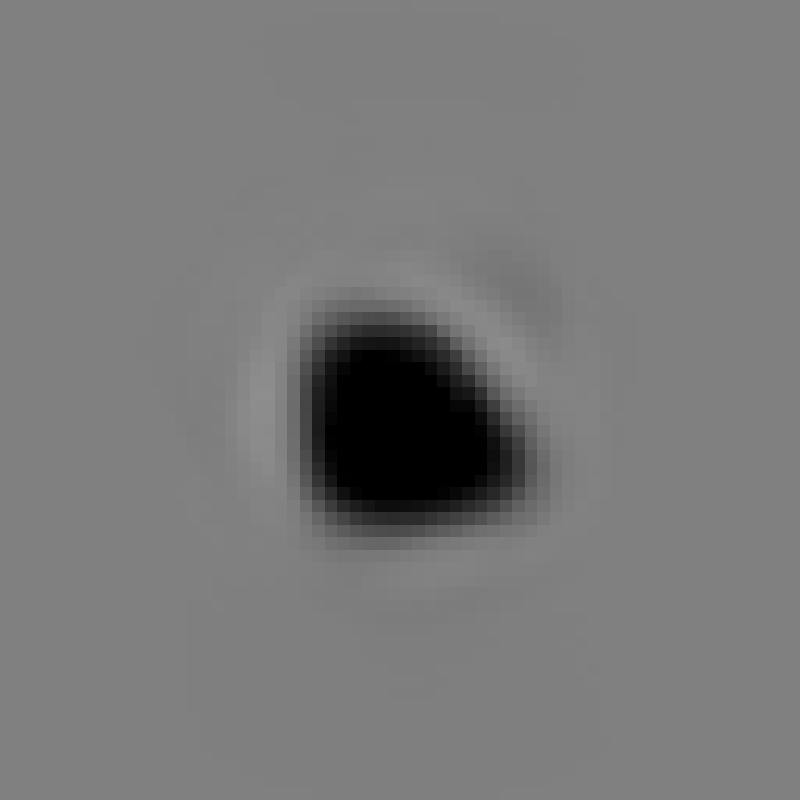}}
\errors{40.1}{19.2}
\colorbarh{-3}{3}
\caption{Exact, retrieved, and adjusted phases, respectively, for an example from the noise-free test set.}
\label{fig:example-test-set2}
\end{figure}

\begin{figure}[htbp]
\centering
{\scalebarbackground{\includegraphics[width=0.3\linewidth]{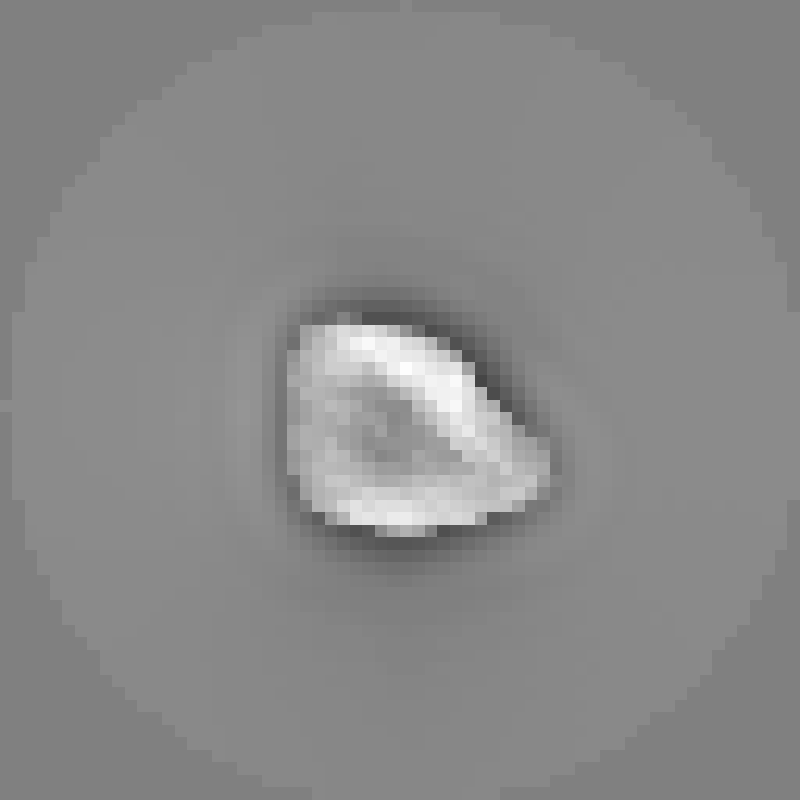}}{150}{1}{50}}{\includegraphics[width=0.3\linewidth]{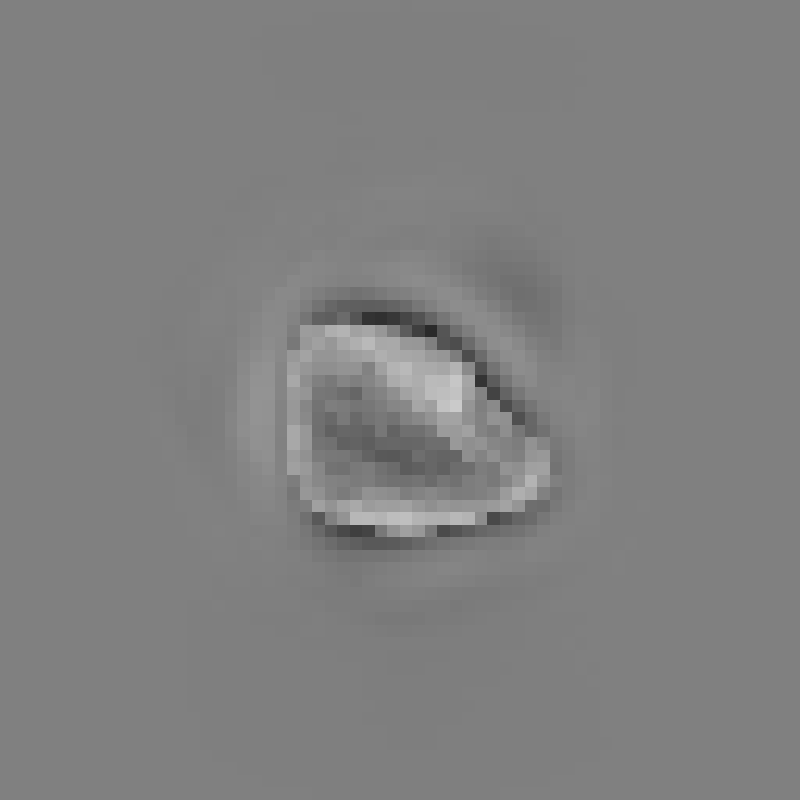}}
\colorbar{-2}{2}
\caption{Errors obtained by applying Eq.~\protect\eqref{eq:error} to the retrieved and adjusted phases from Fig.~\protect\ref{fig:example-test-set2}, respectively.}
\label{fig:example-test-set2-errors}
\end{figure}

The ANN was then trained on the noise-free training set, and we used this to process a test set of $100$ retrieved noise-free phases. The mean rms error was reduced from $\bar E_\textup{rms} = 38.5\%$ to $\bar E_\textup{rms} = 17.5\%$ for this set. An example test pair from this set, along with the adjusted phase, is shown in Fig.~\ref{fig:example-test-set2}. Because the error here is small, and difficult to discern from the differences in the total phases, we also provide the errors themselves, with the gray scale range scaled accordingly. These are shown in Fig.~\ref{fig:example-test-set2-errors}.

A large component of the total error in the retrieved phase is due to a constant offset. To determine how significant this contribution is, we also ran the ANN using corrected phases for both the training and test data in the noise-free case. The correction was performed by subtracting the mean error from the retrieved phases. The mean rms errors over the test set in this case were $26.7\%$ and $17.1\%$ for the retrieved and adjusted phases, respectively. This demonstrates that the ANN is effective at both removing the offset and ameliorating the defocus-induced errors.

The results presented here provide a powerful means for error reduction in propagation based phase retrieval using noisy images, and show significant promise in the noise-free case. Training the ANN simply requires a sufficient volume of simulated training data, which can be generated based on knowledge of experimental parameters, such as defocus and noise level. Unknown parameters (e.g., refractive index/mean inner potential) can be randomly varied within reasonable ranges in the training data. This will increase the required number of training examples, and probably the complexity of the ANN, but should not require altering the process in any fundamental way. In this work, the choice of hyperparameters (e.g., number of layers, number of nodes, and choice of activation function) were mostly arbitrary. Nonetheless, the results presented here are extremely promising, suggesting that more rigorous tuning of hyperparameters could reduce errors further. This can be achieved using manual or algorithmic optimization methods, training the ANN multiple times using different sets of hyperparameters, and minimizing the error on a validation set~\cite{bergstra2011algorithms}. 

These results show that ANNs can be effective at reducing errors in phase retrieval due to both random noise and systematic errors---the latter comprising a range of errors inherent in the numerical implementation of the TIE, such as truncation errors arising from the use of Eq.~\eqref{eq:fda}, sampling errors, and deviations from Eq.~\eqref{eq:tie_num} due to the use of regularization. The error in the retrieved phase using the TIE depends strongly on the chosen defocus. The signal-to-noise ratio (SNR) increases as defocus increases, but this comes at the cost of diffraction-induced blurring~\cite{paganin2004quantitative}. Thus, the choice of defocus is typically a compromise between these two effects. Although we do not address, in this work, how these two broad categories of errors in the adjusted phases vary with defocus, it is likely that the effectiveness of this method is different for the two respective types of error. This implies that the optimal defocus (and, more generally, the choice of phase retrieval algorithm) for processing retrieved phases using an ANN would, in general, be different to that for the retrieved phase itself. That is, if this method performs best on stochastic errors, a very small defocus should be chosen, decreasing the SNR, but reducing the systematic errors that are more difficult to remove using the ANN. Conversely, if the systematic errors are more effectively removed, a much larger defocus should be chosen, resulting in greater defocus-induced errors that can be ameliorated by the ANN, and maximizing the SNR.

The flexibility of this approach suggests that it may be applicable to a variety of error sources that are not addressed here. For example, electron micrographs of crystalline specimens contain delocalization artifacts~\cite{budinger1976measurement,zandbergen1996non}. These artifacts can be avoided through selection of an appropriate objective aperture, or by choosing crystal orientations that minimize these effects. However, in vector tomographic applications~\cite{phatak2010three}, in which at least two orthogonal tilt series (of multiple orientations each) are required, and angle-limiting apertures can be mechanically obstructed at high tilt, these approaches can be impractical. The delocalization artifacts are deterministic, but highly sensitive to crystal orientation~\cite{zandbergen1996non}, and are therefore difficult to precisely predict for arbitrary orientations. This significantly complicates the inverse problem of reconstructing projected potentials from electron micrographs for the purposes of vector tomographic reconstruction, and no proposed solution to this problem has been published. The pattern recognition capabilities of ANNs may be of immense utility in solving such problems. 

Another consideration relating to vector tomographic applications is that the methods described in this work should generalize to higher dimensions and other reconstruction algorithms, enabling them to be applied to reconstructed vector and scalar fields, which can contain significant errors~\cite{kemp2016sources, prabhat20173d}. Note, however, that the large amount of data in a three dimensional (3D) field (and, in particular, a vector field) would require a corresponding increase in the complexity of the neural network, at the very least in terms of the number of nodes in each layer, and this would make such methods computationally infeasible on current standard desktop computing hardware. However, a thorough performance analysis would be required to ascertain whether these methods can be applied to 3D vector fields using specialized computing hardware such as graphics processing unit, or tensor processing unit, clusters.

The use of ANNs in phase retrieval may have much broader implications than those addressed in these simulations. Some preliminary work (not shown here) has demonstrated promise in using the defocused images directly as the input vector for the ANN, bypassing the phase retrieval step entirely. This could provide significant advantage by \emph{avoiding} many of the causes of errors in phase retrieval, rather than ameliorating them \textit{post hoc}. This also allows the ANN to perform any necessary alignment of the micrographs---which can be required due to shifts, in-plane rotations, and scaling of the images---negating the need for implementing image registration algorithms and potentially incurring additional errors. An avenue to explore in future is the use of convolutional neural networks (CNNs), which is the type of network used by Rivenson \textit{et al.}~\cite{rivenson2017phase}, in conjuction with such methods. A CNN employs additional layers which apply convolutions to the input data to detect image features. CNNs are widely used in computer vision applications such as autonomous driving~\cite{john2015pedestrian} and handwritten character recognition~\cite{ciresan2011convolutional,ciregan2012multi}. Due to the convolutional nature of the numerical implementation of TIE (Eq. (\ref{eq:tie_num})), and the excellent pattern recognition capabilities of CNNs, these may have significant benefits over the type of ANN used in the present work, as a means of extracting phase information directly from defocused intensity measurements. 

\vspace{-0.15cm}
\begin{acknowledgments}
\vspace{-0.15cm}
The author would like to thank David Paganin and Tim Petersen for advice, encouragement, and help with preparing the manuscript.
\end{acknowledgments}

\end{document}